\documentclass[prb,aps,superscriptaddress,twocolumn,showpacs,showkeys, english]{revtex4}
\usepackage[T1]{fontenc}
\usepackage[latin9]{inputenc}
\usepackage{float}
\usepackage{amssymb}
\usepackage{amsmath}
\usepackage{graphicx}
\usepackage{babel}
\usepackage{color}
\usepackage{bm}
\usepackage[colorlinks,citecolor=blue]{hyperref}

\begin{document}

\title{A Non-Gaussian Variational Approach to Fermi Polarons in One- and Two-dimensional Lattices}

\author{Ruijin Liu}
\affiliation{Department of Physics, Renmin University of China, Beijing 100872, China}
\author{Yue-Ran Shi}
\affiliation{Department of Physics, Renmin University of China, Beijing 100872, China}
\author{Wei Zhang}
\email{wzhangl@ruc.edu.cn}
\affiliation{Department of Physics, Renmin University of China, Beijing 100872, China}
\affiliation{Beijing Key Laboratory of Opto-electronic Functional Materials and Micro-nano Devices, Renmin University of China,
Beijing 100872, China}

\begin{abstract}
We study the Fermi polaron problem of one mobile spin-up impurity immersed atop the bath consisting of spin-down fermions in one- and two-dimensional square lattices. We solve this problem by applying a variational approach with non-Gaussian states after separating the impurity and the background by the Lee-Low-Pines transformation. The ground state for a fixed total momentum can be obtained via imaginary time evolution for the variational parameters. For the one-dimensional case, the variational results are compared with numerical solutions of the matrix product state method with excellent agreement. In two-dimensional lattices, we focus on the dilute limit, and find a polaron--molecule evolution in consistence with previous results obtained by variational and quantum Monte Carlo methods for models in continuum space. Comparing to previous works, our method provides the lowest ground state energy in the entire parameter region considered, and has an apparent advantage as it does not need to assume {\it in priori} any specific form of the variational wave function. 
\end{abstract}



\maketitle

\section{Introduction}
Polaron is defined as a dressed state formed by a mobile impurity interacting with a medium. Being first proposed by Landau~\cite{Landau} and Pekar~\cite{Pekar} more than half a century ago to describe the dressing effect of an impurity by the elementary excitations of the medium, the concept of polaron has attracted great attention and become a fundamental problem in condensed matter physics, mainly because it plays an essential role in the building block for understanding more complex many-body phenomena~\cite{Landau2}. Depending on whether the host particle excitations obey Bose or Fermi statistics, a polaron can be classified as a Bose polaron or a Fermi polaron. As the Bose polaron was extensively studied in the context of electron-phonon systems, a Fermi polaron is suggested to behave quite differently, since the impurity may undergo a polaron--molecule transition and effectively change its statistics by binding fermions from the background.

In recent years, there has been a significant amount of theoretical work aimed at understanding polaron problems.
This problem has been studied using a variety of tools, such as the variational approach~\cite{Tempere} based on Feynman path-integral formalism~\cite{Feynman}, numerical simulation based upon the diagrammatic quantum Monte Carlo method~\cite{Svistunov,Mishchenko,Kroiss,Vlietinck}, and systematic perturbation expansion~\cite{Rath,Christensen} with the use of the $T$-matrix~\cite{Fetter,Massignan1,Massignan2}. Chevy has provided an instructive variational wave function~\cite{Chevy} that captures the essential properties of the polaron, even on a quantitative level when compared with Monte Carlo calculations. This method can be improved by including more particle-hole pair excitations~\cite{Bruun,Parish1,Parish2}. Specifically, by including one and two particle-hole pairs in the variational ansatz, a polaron--molecule transition in a two-dimensional (2D) Fermi gas was obtained~\cite{Parish1,Parish2}, which compares well with the experimental results~\cite{Koschorreck}.

Ultracold atomic gases with high controllability provide us a particularly clean and flexible platform to explore polaron physics. For example, by making use of the Feshbach resonance in optical or magnetic traps, polaron properties may be studied to great precision across a broad interaction regime from attractive to repulsive interaction in different dimensions. A Fermi polaron was experimentally observed and investigated in highly polarized two-component Fermi gases~\cite{Zwierlein1,Zwierlein2,Partridge,Schirotzek,Kohstall}. The observation of Bose polarons has been reported by radio frequency spectroscopy of ultracold bosonic $^{39}$K atoms~\cite{PhysRevLett.117.055302} and for $^{40}$K impurities in an ultracold atomic gas of $^{87}$Rb~\cite{PhysRevLett.117.055301}. Besides, polarons in a 2D Fermi gas involving spin-orbit coupling was theoretically studied~\cite{Yi}, which may give rise to a novel Fulde-Ferrell-Larkin-Ovchinnikov-like molecular state. Polaron problems in alkaline-earth(-like) atoms with orbital Feshbach resonance~\cite{Zhang} were also discussed theoretically~\cite{Chen,Deng,Chen2}.

In this paper, we consider a highly polarized Fermi Hubbard model with a single spin-up fermion acting as an impurity interacting with a bath consisting of spin-down fermions. We use the non-Gaussian variational method~\cite{Shi}, which for our case can be understood as a combination of the Lee-Low-Pines (LLP) transformation~\cite{Lee} and the Gaussian state approximation, to determine the ground state of the system. Specifically, under the LLP transformation, the impurity degree of freedom can be eliminated and we can obtain a transformed Hamiltonian describing a single component system with host spin-down fermions only. Then we use a Gaussian wave function to approximate the transformed ground state and determine the corresponding variational parameters by imaginary time evolution. We benchmark our results by comparing to the matrix product state (MPS) method~\cite{Schollwock} for 1D lattices. For 2D case, we focus on the dilute limit, which is closely related to continuum systems. By varying the interaction strength, we find a fairly broad region for the system to evolve from polaron to molecule states. The region for the evolution is consistence with the results obtained by Chevy-type variational ansatz~\cite{Parish2}, diagrammatic Monte Carlo simulation~\cite{Vlietinck,Kroiss}, and impurity lattice Monte Carlo method~\cite{Bour}. We emphasize that our approach offers the lowest ground state energy within the entire region of interaction strength considered, and does not require any knowledge about the wave function ansatz, nor any expensive numerical efforts. Besides, as our method does not rely on the dimensionality or specific form of the lattice, it can be straightforwardly generalized to other lattice configurations in various dimensions.

The remainder of this manuscript is organized as follows. In Sec.~\ref{sec:LLP}, we present the polaron problem under consideration and employ the LLP transformation to decouple the impurity degree of freedom from the background. By assuming a Gaussian state as the trial wave function for the transformed single-component Hamiltonian, the ground state of the original model takes the form of a non-Gaussian state by adding back the impurity degree of freedom and reversing the LLP transformation, as discussed in Sec.~\ref{sec:NGS}. A numerical minimization of energy is then applied to find the approximate eigenstate for a given total momentum. In Sec.~\ref{sec:1D}, we study a 1D lattice and benchmark the outcome of non-Gaussian variational approach by the MPS algorithm, while the results for a 2D square lattice in the dilute limit is discussed in Sec.~\ref{sec:2D}. Finally, we summarize the main conclusion in Sec.~\ref{sec:conclusion}.

\section{Hamiltonian and Lee-Low-Pines Transformation}
\label{sec:LLP}

We consider a Fermi Hubbard model for a two-component Fermi system interacting via an on-site interaction on a one-dimensional chain or a two-dimensional square lattice. The lattice spacing $a=1$ is taken as the length unit throughout this manuscript. The Hamiltonian reads
\begin{eqnarray}
H&=&-t\sum_{\langle \mathbf{ij}\rangle,\sigma}c_{\mathbf{i}\sigma}^{\dag} c_{\mathbf{j}\sigma}+g\sum_{\mathbf{i}}n_{\mathbf{i}\uparrow}n_{\mathbf{i}\downarrow}
-\mu\sum_{\mathbf{i}}c_{\mathbf{i}\downarrow}^{\dag}c_{\mathbf{i}\downarrow},
\label{eqn:H}
\end{eqnarray}
where $c_{\mathbf{i}\sigma}^{\dag}$ and $c_{\mathbf{i}\sigma}$ stand for creation and annihilation operators for fermions on site $\mathbf{i}$ with spin $\sigma = \uparrow,\downarrow$, $n_{i\sigma}=c_{\mathbf{i}\sigma}^{\dag} c_{\mathbf{i}\sigma}$ is the number operator, $\mu$ is the chemical potential to tune the number of spin-down particles, and the summation in the first term runs over all nearest neighboring sites $\langle \mathbf{ij}\rangle$. To study the polaron physics, we focus on the highly polarized limit with only one single spin-up impurity, i.e., $N_{\uparrow}=\sum\limits_{\mathbf{i}}n_{\mathbf{i}\uparrow}=1$.

Notice that the system possesses translational symmetry and the total momentum is a good quantum number. To eliminates the impurity degree of freedom, we introduce a unitary transformation
\begin{eqnarray}
\label{eqn:LLP}
U_{\rm LLP}=e^{-i\mathbf{Q}\cdot\mathbf{X}},
\end{eqnarray}
where $\mathbf{Q}=\sum\limits_{\mathbf{k}} \mathbf{k}c_{\mathbf{k}\downarrow}^\dag c_{\mathbf{k}\downarrow}$ is the total momentum operator of the spin-down background, $\mathbf{k}$ is the reciprocal lattice vector, and $\mathbf{X}=\sum\limits_\mathbf{i} \mathbf{i}c_{\mathbf{i}\uparrow}^{\dag}c_{\mathbf{i}\uparrow}$ is the coordinate operator of the spin-up impurity. The transformation Eq.~(\ref{eqn:LLP}), known as the Lee-Low-Pines (LLP) transformation, is introduced in 1953 to study the problem of an impurity fermion immersed in a background of phonons~\cite{Lee}. In the following discussion, we show the same transformation can separate the degrees of freedom of the spin-up impurity and the spin-down Fermi sea, as it does in a Bose medium of phonons.

We first rewrite the spin-down part of the Hamiltonian Eq.~(\ref{eqn:H}) in momentum space
\begin{eqnarray}
\label{eqn:H2}
H=&-&t\sum_{\langle \mathbf{ij}\rangle}c_{\mathbf{i}\uparrow}^{\dag} c_{\mathbf{j}\uparrow}+\sum_{\mathbf{k}}(\varepsilon_{\mathbf{k}}-\mu)c_{\mathbf{k}\downarrow}^\dag c_{\mathbf{k}\downarrow}\nonumber\\
&+&\frac{g}{\Omega}\sum_{\mathbf{i},\mathbf{k},\mathbf{k}'}c_{\mathbf{i}\uparrow}^{\dag} c_{\mathbf{i}\uparrow}e^{i(\mathbf{k}'-\mathbf{k})\cdot\mathbf{i}}c_{\mathbf{k}\downarrow}^{\dag}c_{\mathbf{k}'\downarrow},
\end{eqnarray}
where $c_{\mathbf{k}\downarrow}=\frac{1}{\sqrt{\Omega}}\sum\limits_{\mathbf{i}} e^{-i\mathbf{k}\cdot\mathbf{i}}c_{\mathbf{i}\downarrow}$ and
$c_{\mathbf{k}\downarrow}^\dagger=\frac{1}{\sqrt{\Omega}}\sum\limits_{\mathbf{i}} e^{i\mathbf{k}\cdot\mathbf{i}}c_{\mathbf{i}\downarrow}^\dagger$ are the fermion operators in momentum with $\Omega$ the number of lattice sites, and the dispersion reads $\varepsilon_{\mathbf{k}}=-2t\cos k$ and $\varepsilon_{\mathbf{k}}=-2t(\cos k_x+\cos k_y)$ for 1D and 2D lattices, respectively. Next, we apply the LLP transformation on Eq.~(\ref{eqn:H2}). Using the Baker-Campbell-Hausdorff (BCH) formula, the fermion operators $c_{\mathbf{k}\downarrow}$ and $c_{\mathbf{i}\uparrow}$ transform as $U_{\rm LLP}^\dag c_{\mathbf{k}\downarrow} U_{\rm LLP}=e^{-i\mathbf{k}\cdot\mathbf{X}}c_{\mathbf{k}\downarrow}$ and $U_{\rm LLP}^\dag c_{\mathbf{i}\uparrow} U_{\rm LLP}=e^{-i\mathbf{Q}\cdot\mathbf{i}}c_{\mathbf{i}\uparrow}$, respectively. The Hamiltonian after the LLP transformation then takes the following form
\begin{eqnarray}
&&H_{\rm LLP}=\sum_{\mathbf{k}}c_{\mathbf{k}\uparrow}^{\dag} c_{\mathbf{k}\uparrow}\left[-t\sum_{ \bm{\delta}}e^{-{i(\mathbf{k}-\mathbf{Q})\cdot\bm{\delta}}}\right]
\nonumber \\
&&+\sum_{\mathbf{k}}(\varepsilon_{\mathbf{k}}-\mu)c_{\mathbf{k}\downarrow}^\dag c_{\mathbf{k}\downarrow}
+\sum_{\mathbf{k}'}c_{\mathbf{k}'\uparrow}^{\dag} c_{\mathbf{k}'\uparrow}\left[\frac{g}{\Omega}\sum_{\mathbf{k},\mathbf{q}}c_{\mathbf{k}\downarrow}^{\dag}c_{\mathbf{q}\downarrow}\right].
\label{eqn:HLLP}
\end{eqnarray}
Here, $\bm{\delta}$ are the unit lattice vectors with $\bm{\delta}= \pm 1$ for 1D and $\bm{\delta} \in \{(\pm 1,0), (0, \pm 1)\}$ for 2D lattices. It can be seen that the LLP transformation separates explicitly the total conserved momentum of the system. Indeed, the total conserved momentum is transformed as the momentum of spin-up particle, which, for a given total momentum, eliminates the degree of freedom of the impurity. Thus, for a given total momentum $\mathbf{K}$, the problem reduces to a Hamiltonian containing spin-down component only
\begin{eqnarray}
H_{\mathbf{K}\downarrow}=&&\sum_{\mathbf{k}}(\varepsilon_{\mathbf{k}}-\mu)c_{\mathbf{k}\downarrow}^\dag c_{\mathbf{k}\downarrow}+\frac{g}{\Omega}\sum_{\mathbf{k},\mathbf{q}}c_{\mathbf{k}\downarrow}^{\dag}c_{\mathbf{q}\downarrow}\nonumber\\
&&
-t\sum_{\bm{\delta}}e^{-{i(\mathbf{K}-\mathbf{Q}) \cdot \bm{\delta}}}.
\label{eqn:Hdown}
\end{eqnarray}
Next, we construct a variational wave function in a Gaussian form to find the approximate ground state of the Hamiltonian above, and then obtain the eigenstate of the original Hamiltonian with a total conserved momentum by adding back the spin-up impurity and reversing the LLP transformation.

\section{Non-Gaussian state variational approach}
\label{sec:NGS}

The essence of the variational approach used in this work is to approximate the ground state of Eq.~(\ref{eqn:Hdown}) by a Gaussian trial wave function
\begin{eqnarray}
|\Psi_{\rm GS}\rangle=U_{\rm GS}|0\rangle_{\downarrow},
\end{eqnarray}
where the unitary transformation takes the form $U_{\rm GS}=e^{i\frac{1}{4}A^{T}\xi_{m}A}$ with $A=(a_{1,\mathbf{k}_1},\ldots,a_{1,\mathbf{k}_{\Omega}},a_{2,\mathbf{k}_1},\ldots,a_{2,\mathbf{k}_{\Omega}})^{T}$. The Majorana operators for spin-down fermions
are defined as $a_{1,\mathbf{k}_j}=c_{\mathbf{k}_j,\downarrow}^{\dag}+c_{\mathbf{k}_j,\downarrow}$ and
$a_{2,\mathbf{k}_j}=i(c_{\mathbf{k}_j,\downarrow}^{\dag}-c_{\mathbf{k}_j,\downarrow})$,
and satisfy the anti-commutation relation
$\{a_{\alpha,\mathbf{k}},a_{\beta,\mathbf{k}'}\}=2\delta_{\alpha\beta}\delta_{\mathbf{k}\mathbf{k'}}$.
The variational parameter $\xi_m$ is an antisymmetric Hermitian matrix. To eliminate the gauge degree of freedom in $\xi_m$, it is convenient to introduce a covariance matrix~\cite{Shi}
\begin{eqnarray}
(\Gamma_m)_{i,j}=\frac{i}{2}\langle\Psi_{\rm GS}|[A_i,A_j]|\Psi_{\rm GS}\rangle,
\label{eqn:gammam}
\end{eqnarray}
where $A_i$ labels the $i$-th element of $A$. The covariance matrix is related to $\xi_m$ as
\begin{eqnarray}
\Gamma_m=-U_{m}\Sigma U_{m}^{T},
\label{eqn:covariant}
\end{eqnarray}
where $U_m=e^{i\xi_m}$, and $\Sigma$ is constructed by the identity matrix ${\openone}_{\Omega}$ of dimension $\Omega$ as
\begin{equation}
\label{eqn:Sigma}
\Sigma \equiv i\sigma_y\otimes {\openone}_{\Omega}
= \left(
\begin{array}{cc}
0 & {\openone}_{\Omega} \\
-{\openone}_{\Omega} & 0
\end{array}
\right) .
\end{equation}

By adding back the spin-up impurity and reversing the LLP transformation, the eigenstate of the original Hamiltonian Eq.~(\ref{eqn:H}) with a total conserved momentum $\mathbf{K}$ can be expressed as a non-Gaussian state
\begin{eqnarray}
|\Psi_{\rm NGS}\rangle=U_{\rm LLP}(c_{\mathbf{K}\uparrow}^{\dag}\otimes U_{\rm GS})|0\rangle.
\label{eqn:NGSstate}
\end{eqnarray}
It can be seen that this variational ansatz contains dressing effect of an arbitrary number of particle-hole excitations atop the spin-down Fermi sea, as can be seen in a series expansion of the exponential function of the Gaussian state.

In the spirit of variational method, the ground state of a Hamiltonian $H$ can be obtained via an imaginary time evolution of a trial wave function
\begin{eqnarray}
|\Psi(\tau)\rangle&=&\frac{e^{-H\tau}|\Psi(0)\rangle}{\sqrt{\langle\Psi(0)|e^{-2H\tau}|\Psi(0)\rangle}}
\label{eqn:ima1}
\end{eqnarray}
to the asymptotic limit $\tau\rightarrow\infty$ provided that the initial trial state $|\Psi(0)\rangle$ has a nonzero overlap with the ground state.
Such an evolution can be described by an differential equation
\begin{eqnarray}
d_{\tau}|\Psi(\tau)\rangle&=&-(H-\langle H\rangle)|\Psi(\tau)\rangle
\label{eqn:ima2}
\end{eqnarray}
with the mean energy $\langle H\rangle=\langle\Psi(\tau)|H|\Psi(\tau)\rangle$.
Thus, the imaginary-time evolution equation for the non-Gaussian state Eq.~(\ref{eqn:NGSstate}) can be written as
\begin{eqnarray}
d_{\tau}|\Psi_{\rm NGS}\rangle&=&-\mathcal{P}(H-E)|\Psi_{\rm NGS}\rangle,\label{ima3}
\end{eqnarray}
where $E=\langle\Psi_{\rm NGS}|H|\Psi_{\rm NGS}\rangle=\langle\Psi_{\rm GS}|H_{\mathbf{K}\downarrow}|\Psi_{\rm GS}\rangle$ is the variational mean energy and $\mathcal{P}$ is the projection operator onto the subspace spanned by tangent vectors of the variational manifold. The left-hand side of Eq.~(\ref{ima3}) gives
\begin{eqnarray}
d_{\tau}|\Psi_{\rm NGS}\rangle=U_{\rm LLP}[(c_{\mathbf{K}\uparrow}^{\dagger}|0\rangle_{\uparrow})\otimes (U_{\rm GS}U_{L}|0\rangle_{\downarrow})],
\label{eqn:left}
\end{eqnarray}
where
\begin{eqnarray}
U_{L}=\frac{1}{4}\text{:}A^{T}U_{m}^{T}(\partial _{\tau }U_{m})A\text{:}+\frac{i}{4}{\rm Tr}\left[U_{m}^{T}(\partial _{\tau }U_{m})\Gamma_{m}\right],
\end{eqnarray}
and $:\ :$ represents normal ordering with respect to the vacuum state.
The right-hand side of Eq.~(\ref{ima3}) reads
\begin{eqnarray}
&&-(H-\langle H\rangle)|\Psi_{\rm NGS}\rangle=
\nonumber \\
&&\hspace{1cm}-U_{\rm LLP}\left[(c_{\mathbf{K}\uparrow}^{\dagger}|0\rangle_{\uparrow})\otimes (U_{\rm GS}U_{R}|0\rangle_{\downarrow})\right]
\label{eqn:right}
\end{eqnarray}
where $U_{R}=(i/4)\text{:}A^{T}U_{m}^{T}h_{m}U_{m}A\text{:}+\delta H_{\mathbf{K}\downarrow}$. Here, $\delta H_{\mathbf{K}\downarrow}$ denotes the higher order terms of $c_{\bm{k}\downarrow}$ that are orthogonal to the tangential
space which will be projected out by $\mathcal{P}$ in Eq.~(\ref{ima3}), and
\begin{eqnarray}
h_{m}=4\frac{\delta E}{\delta\Gamma_{m}}
\label{hm}
\end{eqnarray}
is the functional derivative of the variational energy. Comparing Eqs.~(\ref{eqn:left}) and (\ref{eqn:right}), and combining the covariant parameter defined by Eq.~(\ref{eqn:covariant}), we can finally obtain the imaginary time equation of motion (EOM) for the covariance matrix $\Gamma_m$
\begin{equation}
\partial _{\tau }\Gamma_{m}=-h_{m}-\Gamma_{m}h_{m}\Gamma_{m}.
\label{eqn:EOM}
\end{equation}

To evolve the variational parameter $\Gamma_m$ according to EOM given by Eq.~(\ref{eqn:EOM}), we need to calculate the functional derivative $h_m$ defined in Eq.~(\ref{hm}). First of all, we calculate the variational energy $E=\langle\Psi_{\rm GS}|H_{\mathbf{K}\downarrow}|\Psi_{\rm GS}\rangle$. Using the relations $c_{\mathbf{k},\downarrow}^{\dagger}=\frac{1}{2}(a_{1,\mathbf{k}}-ia_{2,\mathbf{k}})$ and $c_{\mathbf{k},\downarrow}=\frac{1}{2}(a_{1,\mathbf{k}}+ia_{2,\mathbf{k}})$, we can rewrite the first and the second terms of $H_{\mathbf{K}\downarrow}$ in Eq.~(\ref{eqn:Hdown}) as
\begin{widetext}
\begin{eqnarray}
&&\sum_{\mathbf{k}}(\varepsilon_{\mathbf{k}}-\mu)c_{\mathbf{k}\downarrow}^\dag c_{\mathbf{k}\downarrow}+\frac{g}{\Omega}\sum_{\mathbf{k},\mathbf{q}}c_{\mathbf{k}\downarrow}^{\dag}c_{\mathbf{q}\downarrow} \nonumber\\
&&\hspace{0.5cm}=\frac{1}{4}\sum_{\mathbf{k}}(\varepsilon_{\mathbf{k}}-\mu)(a_{1,\mathbf{k}}a_{1,\mathbf{k}}+a_{2,\mathbf{k}}a_{2,\mathbf{k}}-ia_{2,\mathbf{k}}a_{1,\mathbf{k}}+ia_{1,\mathbf{k}}a_{2,\mathbf{k}})
+\frac{g}{4\Omega}\sum_{\mathbf{k},\mathbf{q}}(a_{1,\mathbf{k}}a_{1,\mathbf{q}}+a_{2,\mathbf{k}}a_{2,\mathbf{q}}-ia_{2,\mathbf{k}}a_{1,\mathbf{q}}+ia_{1,\mathbf{k}}a_{2,\mathbf{q}})\nonumber\\
&&\hspace{0.5cm}=\frac{1}{2}\sum_{\mathbf{k}}\left(\varepsilon_{\mathbf{k}}-\mu+\frac{g}{\Omega}\right)+\frac{i}{4}A^{T}H_0A-\frac{i\mu}{4}A^{T}\Sigma A.
\end{eqnarray}
The matrix $\Sigma$ is defined as in Eq.~(\ref{eqn:Sigma}), and $H_0=i\sigma_y\otimes[diag(\varepsilon_{\mathbf{k}} ) +({g}/{\Omega})ones(\Omega)]$, where $diag(\varepsilon_{\mathbf{k}} )$ is an $\Omega\times\Omega$ diagonal matrix with diagonal matrix elements $\varepsilon_{\mathbf{k}_1},\cdots,\varepsilon_{\mathbf{k}_{\Omega}}$,  and $ones(\Omega)$ is an $\Omega\times\Omega$ matrix with all elements being 1.
The expectation value of the term $(i/4)A^{T}H_0A$ under the Gaussian state can be calculated as
\begin{eqnarray}
&&\frac{i}{4}\langle\Psi_{\rm GS}|A^{T}H_0A|\Psi_{\rm GS}\rangle=\frac{i}{4}\sum_{i,j}(H_0)_{i,j}\langle\Psi_{\rm GS}|A_i A_j|\Psi_{\rm GS}\rangle\nonumber\\
&&\hspace{0.5cm}=\frac{i}{4}\sum_{i<j}\left[(H_0)_{i,j}\langle\Psi_{\rm GS}|A_i A_j|\Psi_{\rm GS}\rangle+(H_0)_{j,i}\langle\Psi_{\rm GS}|A_j A_i|\Psi_{\rm GS}\rangle \right]
=\frac{i}{4}\sum_{i<j}(H_0)_{i,j}\langle\Psi_{\rm GS}|([A_i, A_j])|\Psi_{\rm GS}\rangle
\nonumber \\
&&\hspace{0.5cm}=\frac{1}{2}\sum_{i<j}(H_0)_{i,j}(\Gamma_m)_{i,j}
=\frac{1}{4}\sum_{i,j}(H_0)_{i,j}(\Gamma_m)_{i,j},
\label{eqn:mean1}
\end{eqnarray}
\end{widetext}
where we have used the antisymmetry of $H_0$ and the covariance matrix $\Gamma_m$ defined in Eq.~(\ref{eqn:gammam}).
In the same way, we have
\begin{eqnarray}
\frac{i\mu}{4}\langle\Psi_{\rm GS}|A^{T} \Sigma A |\Psi_{\rm GS}\rangle=\frac{\mu}{4}\sum_{i,j} \Sigma_{i,j}(\Gamma_m)_{i,j}.
\label{eqn:mean2}
\end{eqnarray}

The mean value of operators that take the form as the third term in Eq.~(\ref{eqn:Hdown}) can be obtained by introducing coherent representation for the fermionic Gaussian state, and the result is~\cite{Shi}
\begin{eqnarray}
\langle\Psi_{\rm GS}| e^{i\mathbf{Q}\cdot\bm{\delta}}|\Psi_{\rm GS}\rangle
= \left(- \frac{1}{2} \right)^{\Omega}s_{f}\text{Pf}(\Gamma _{F}),
\label{eqn:mean3}
\end{eqnarray}
where $s_{f}=(-1)^{\Omega/2}$ and $s_{f}=(-1)^{(\Omega-1)/2}$ for $\Omega$ being even and odd, respectively. Other quantities in the expression above are $\Gamma_{F}=\sqrt{1-e^{i\alpha }}\Gamma _{m}\sqrt{1-e^{i\alpha }}-(1+e^{i\alpha })\Sigma$,
$\alpha={\openone}_{2}\otimes diag(\mathbf{k}\cdot\bm{\delta})$ with $diag(\mathbf{k}\cdot\bm{\delta})$ a diagonal matrix with diagonal elements $\mathbf{k}_1\cdot\bm{\delta},\cdots,\mathbf{k}_{\Omega}\cdot\bm{\delta}$,
and Pf$(\Gamma _{F})$\ denotes the Pfaffian of $\Gamma _{F}$.
Combining Eqs. (\ref{eqn:mean1}), (\ref{eqn:mean2}) and (\ref{eqn:mean3}), we obtain the variational energy
\begin{eqnarray}
E&=&\langle\Psi_{\rm GS}|H_{\mathbf{K}\downarrow}|\Psi_{\rm GS}\rangle\nonumber\\
&=&\frac{1}{2}\sum_{\mathbf{k}}\varepsilon_{\mathbf{k}}-\frac{\Omega\mu}{2}+\frac{g}{2}+\frac{1}{4}\sum_{i,j}(H_0)_{i,j}(\Gamma_m)_{i,j}\nonumber\\
&&-t\sum_{ \bm{\delta}}e^{-i\mathbf{K}\cdot \bm{\delta}}
\left(-\frac{1}{2}\right)^{\Omega}s_{f}\text{Pf}(\Gamma _{F}) -\frac{\mu}{4}\sum_{i,j}\Sigma_{i,j}(\Gamma_m)_{i,j}.\nonumber\\
\label{eqn:energy}
\end{eqnarray}
The functional derivative $h_m$ is
\begin{eqnarray}
h_{m}&=&H_0-\mu\Sigma+2t\sum_{ \bm{\delta}}\bigg[e^{-i\mathbf{K}\cdot\bm{\delta}} \left(-\frac{1}{2}\right)^{\Omega}
\nonumber \\
&& \times
s_{f}\text{Pf}(\Gamma_F)\sqrt{1- e^{i\alpha }}\Gamma_F^{-1}\sqrt{1- e^{i\alpha }}\bigg].
\label{eqn:hm1}
\end{eqnarray}
In addition, the particle number of the spin-down medium is determined by
\begin{eqnarray}
&&N_{\downarrow}=-\frac{\partial E}{\partial\mu}=\frac{\Omega}{2}+\frac{1}{4}\sum_{i,j}\Sigma_{i,j}(\Gamma_m)_{i,j}.
\label{eqn:num}
\end{eqnarray}
In the following discussion, we evolve the variational parameter $\Gamma_m$ via Eqs.~(\ref{eqn:EOM}) and(\ref{eqn:hm1}) until a convergence of variational energy given by Eq.~(\ref{eqn:energy}) is reached under a number constraint Eq.~(\ref{eqn:num}).

\section{One-dimensional case}
\label{sec:1D}
\begin{figure}[t]
	\begin{center}
		\includegraphics[width=8cm]{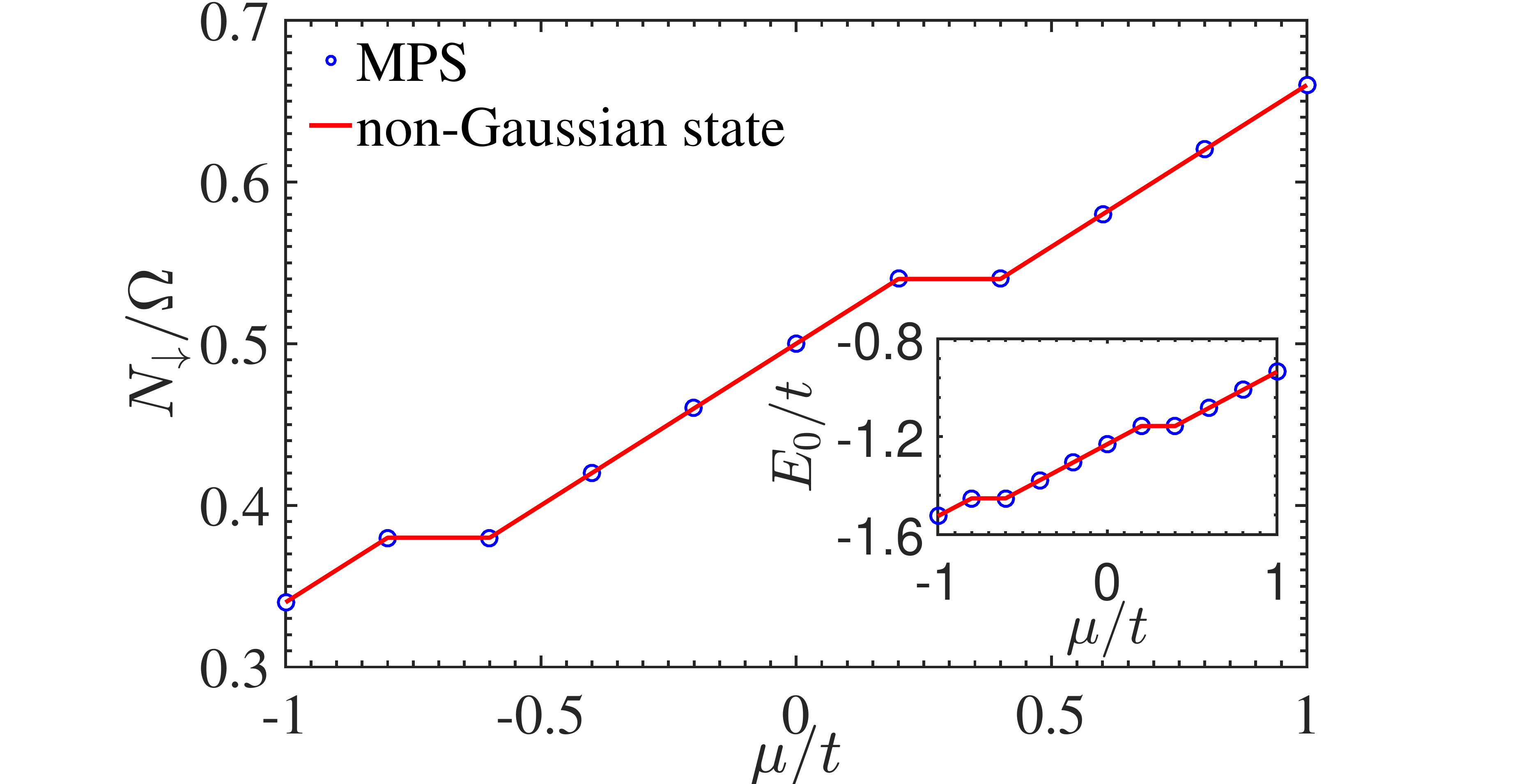}
	\end{center}
	\caption{\label{fig:1dnmu}(Color online) Particle number of the spin-down component and ground state energy of a Fermi polaron (inset) in a 1D lattice by varying the chemical potential. Results obtained by non-Gaussian variational approach and MPS method reach an excellent agreement. Parameters used in this figure are $g/t=2$, $\mathbf{K}=0$ and $\Omega=50$.}
\end{figure}
\begin{figure}
	\begin{center}
		\includegraphics[width=9cm]{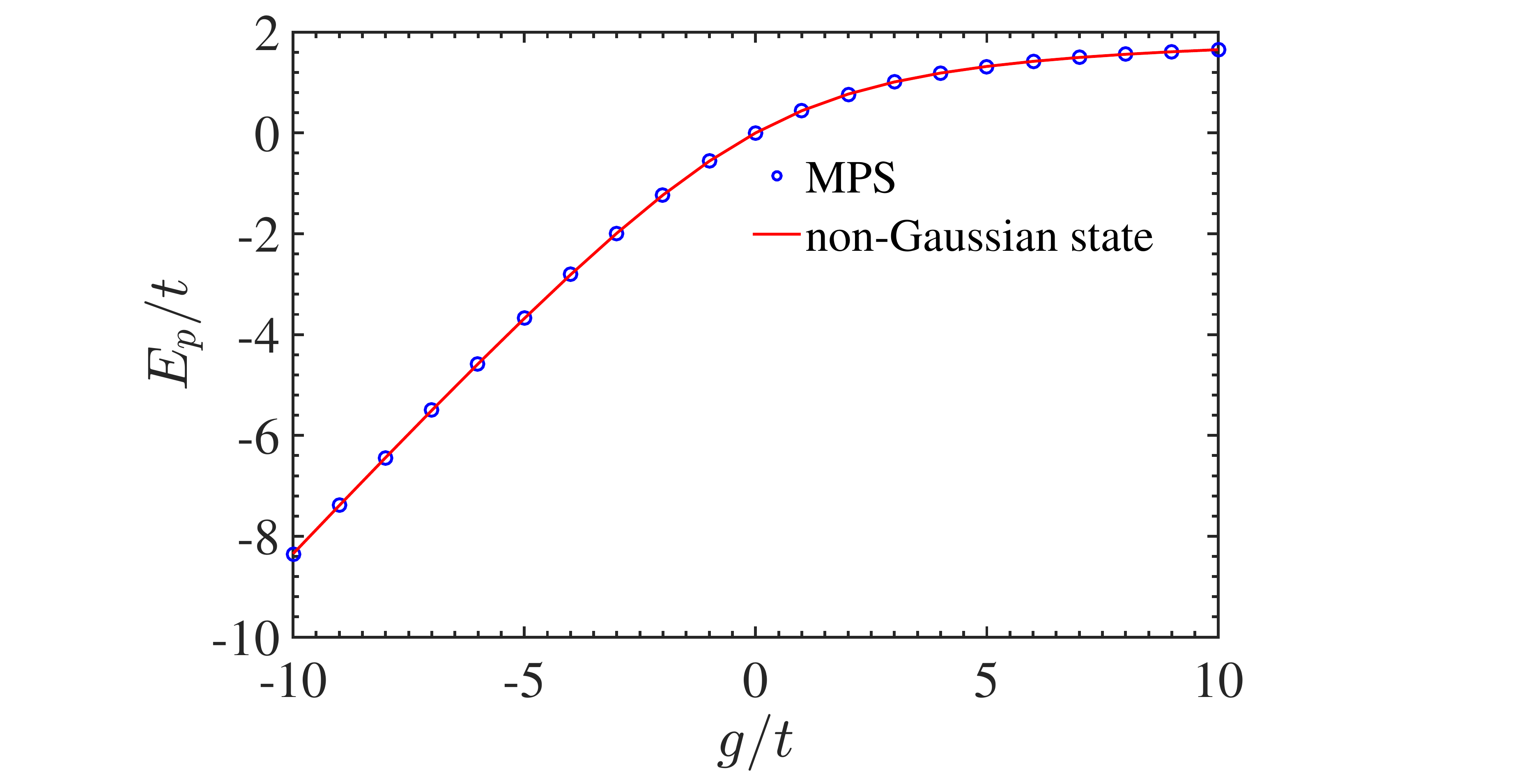}
	\end{center}
	\caption{\label{fig:1dEg}(Color online) Ground state energy of a Fermi polaron in a 1D lattice by varying the interaction strength. Other parameters are taken as $N_{\downarrow}=25$, $\mathbf{K}=0$ and $\Omega=50$.}
\end{figure}

First, we focus on the one-dimensional case and study the dispersion of the system for a given total momentum $\mathbf{K}$. In our numerical variation, we evolve the imaginary time EOM (\ref{eqn:EOM}) until a convergence to a steady state is reached. To ensure the resulting state is the true ground state, we run the evolution for a set of randomly generated initial states and choose the outcome with lowest energy. The variational results are then compared with those obtained by the matrix product states (MPS) algorithm under a periodic boundary condition.

Figures~\ref{fig:1dnmu} and \ref{fig:1dEg} show the results of total momentum $\mathbf{K}=0$. We first point out that the results obtained by non-Gaussian state with $\mathbf{K}=0$ agrees perfectly well with those obtained by MPS method without specifying the total momentum $\mathbf{K}$, indicating that the ground state has a zero total momentum. Owing to the finite size effect, we observe a step-like jump in both the particle number and the ground state energy by varying the chemical potential, as depicted in Fig.~\ref{fig:1dnmu}. To further elucidate the interaction effect, in this figure and the following discussion we set the zero point energy to be the energy of the corresponding non-interacting case, and define the polaron energy as $E_p(\mathbf{K},g)=E(\mathbf{K},g)-E(\mathbf{K}=0,g=0)$. Notice that the non-interacting system energy $E(\mathbf{K}=0,g=0)$ can be calculated exactly.

In Fig.~\ref{fig:1dEg}, we fix the density of spin-down particles $N_{\downarrow}/\Omega=0.5$ and vary the interaction $g$ from attractive to repulsive. It can be seen that the ground state energy varies smoothly versus interaction. In the limit of infinitely large repulsion $g \to +\infty$, the spin-up impurity acts an effective hard wall for spin-down particles, which cuts two links with hopping rate $t$ and hence leads to an energy $E_p \to 2t$. On the contrary limit of large attractive interaction $g \to -\infty$, the spin-up impurity is tightly bound with one spin-down particle, and working together as an impenetrable boundary due to the Pauli blocking effect. Thus, the energy tends to the limiting value of $E_p \to g + 2t$. Our numerical results are consistent with the two limits.
\begin{figure}
	\begin{center}
		\includegraphics[width=9cm]{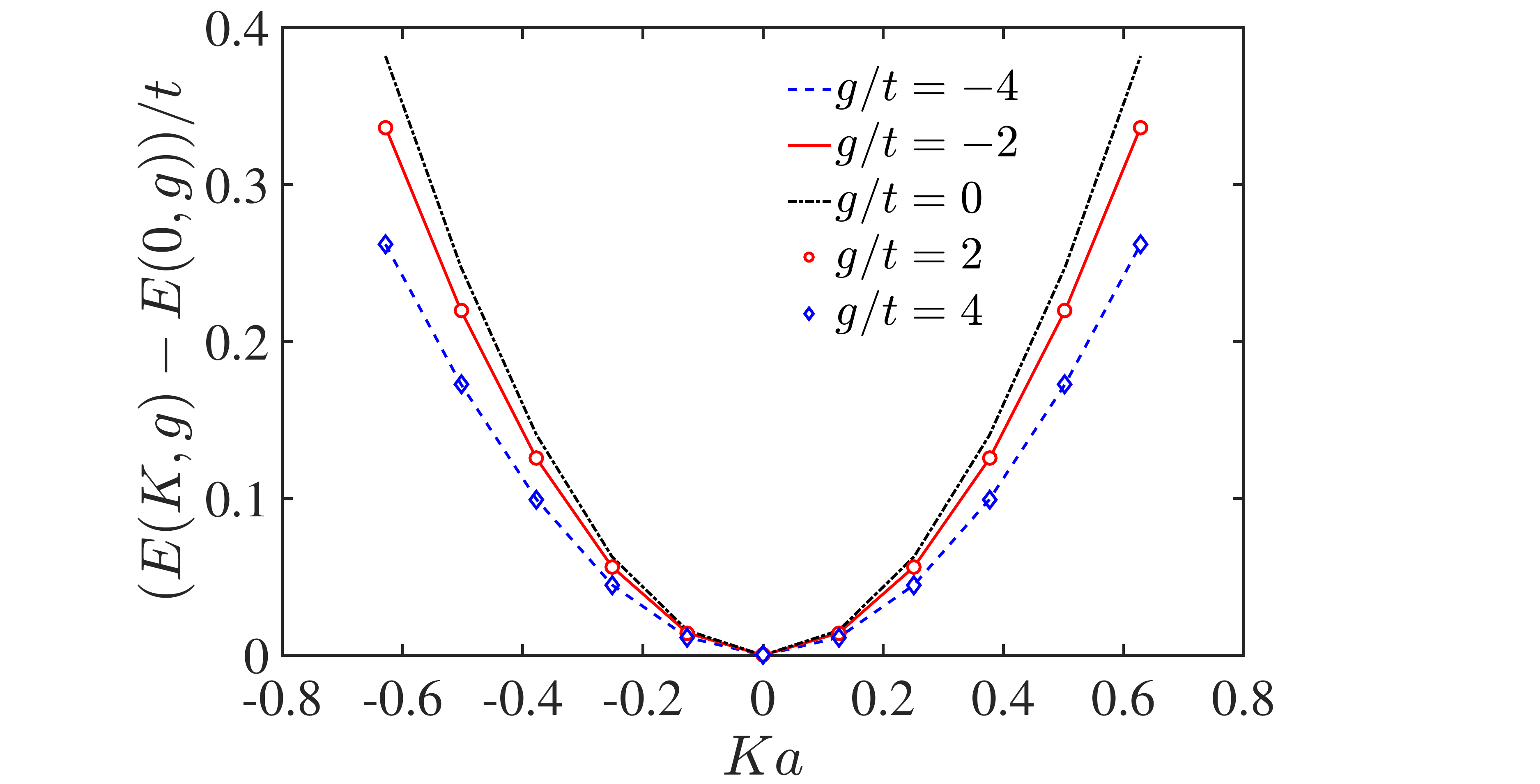}
	\end{center}
	\caption{\label{fig:1dEk}(Color online) Dispersion relations for a 1D Fermi polaron with various interaction strengths. Results for interaction with opposite signs are identical owing to the partial particle-hole symmetry as discussed in the main text. The effective mass increases with interaction strength $|g|$, revealing a more significant dressing effect induced by the bath. Other parameters used in this plot are $N_{\downarrow}=25$ and $\Omega=50$.}
\end{figure}

Next, we fix the density of spin-down particles at half filling with $\mu = 0$ and $N_{\downarrow}/\Omega=0.5$, and extract the dispersion relation $E(\mathbf{K},g)-E(0,g)$ by varying the total momentum $\mathbf{K}$ with interaction strength $g/t=0, \pm 2,\pm 4$. From Fig.~\ref{fig:1dEk}, we find that the effective mass of the quasiparticle defined as
\begin{equation}
m^{*}=\left(\frac{\partial^2E}{\partial K^2}\bigg|_{K=0}\right)^{-1}
\end{equation}
is independent of the sign of interaction and increases monotonically with $|g|$. The symmetry respected to the sign of interaction can be understood by applying a partial particle-hole transformation~\cite{Tian}
\begin{eqnarray}
U_{\downarrow}\equiv\prod_{j} \left[c_{j\downarrow}+(-1)^{j}c_{j\downarrow}^{\dag} \right],
\label{eqn:pht}
\end{eqnarray}
which transforms the operators $c_{i\sigma}$ as
\begin{eqnarray}
U_{\downarrow}^{\dag}c_{i\uparrow}U_{\downarrow}=c_{i\uparrow},\quad\quad U_{\downarrow}^{\dag}c_{i\downarrow}U_{\downarrow}=(-1)^ic_{i\downarrow}^{\dag}.
\end{eqnarray}
At half filling with $\mu=0$, the Hamiltonian Eq.~(\ref{eqn:H}) is transformed under Eq.~(\ref{eqn:pht}) as
\begin{eqnarray}
U_{\downarrow}^{\dag}H(g,\mu=0)U_{\downarrow}=H(-g,\mu=0)+g\sum_{i}c_{i\uparrow}^{\dag}c_{i\uparrow},
\end{eqnarray}
while the number constraint remains unchanged, i.e., $N_{\uparrow}=1$ and $N_{\downarrow}/\Omega=0.5$. Thus we have the relation of energy spectra for interaction of opposite signs
\begin{eqnarray}
E(\mathbf{K},g)=E(\mathbf{K},-g)+g
\label{eqn:shiftE}
\end{eqnarray}
at half filling. This result shows that the difference between $E(\mathbf{K},g)$ and $E(\mathbf{K},-g)$ is a constant $g$, while the effective masses for the two cases are equivalent.

\section{Two-dimensional case in the dilute limit}
\label{sec:2D}
\begin{figure}
	\begin{center}
		\includegraphics[width=9cm]{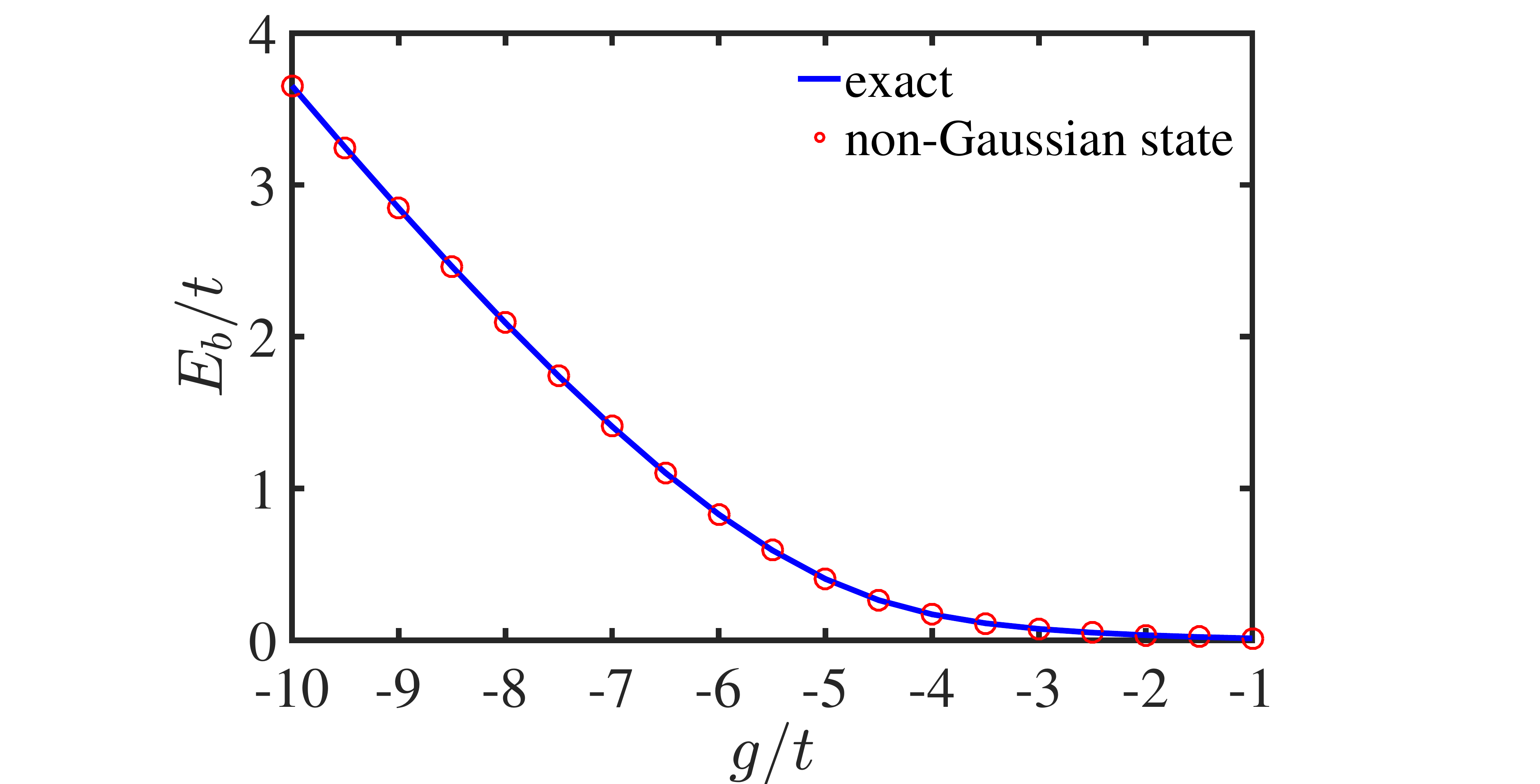}
	\end{center}
	\caption{\label{fig:twobody}(Color online) Two-body bound state energy for a single spin-up and a single spin-down atoms in a 2D square lattice of size $50 \times 50$. The exact result is obtained by solving the two-body problem analytically as in Eq.~(\ref{eqn:2body}).}
\end{figure}

In this section, we employ the non-Gaussian variational approach to a 2D square lattice. We focus on the dilute limit with the number of spin-down particles is much smaller than that of lattice sites, i.e., $N_\downarrow \ll \Omega$. This limit is of particular interest as it is closely related to the continuum model, which can be considered as a lattice model with an infinitesimal lattice spacing $d \to 0$. The problem of Fermi polaron in a 2D continuum system has been studied by various methods.~\cite{Parish1,Parish2,Vlietinck,Kroiss,Bour} Previous works using a variational approach by including more pairs of particle-hole excitations show that there exists a polaron--molecule transition in the ground state as the interaction varies.~\cite{Parish1,Parish2} Similar findings have been obtained in diagrammatic Monte Carlo (diagMC) simulations.~\cite{Vlietinck,Kroiss} All these variational and diagMC studies perform separate calculations for polaron and molecule states, where the transition is identified as the level crossing point of the two states. Later, in order to study the transition region in a unified way, a fully non-perturbative calculation was performed using the impurity lattice Monte Carlo (ILMC) method.~\cite{Bour} One feature of the ILMC method is the discretization of the spatial part. The results obtained by ILMC shows evidence for a smooth crossover from polaron to molecule states. Here, we study the 2D lattice model in the dilute limit via the non-Gaussian variational approach, without assuming {\it in priori} any specific form of the wave function. In the following calculation, we take the lattice size as $\Omega=50\times50$ and $N_{\downarrow} \approx 37$, which corresponds to a filling density $N_\downarrow / \Omega \approx 0.015$.


\begin{figure}
	\begin{center}
		\includegraphics[width=8cm]{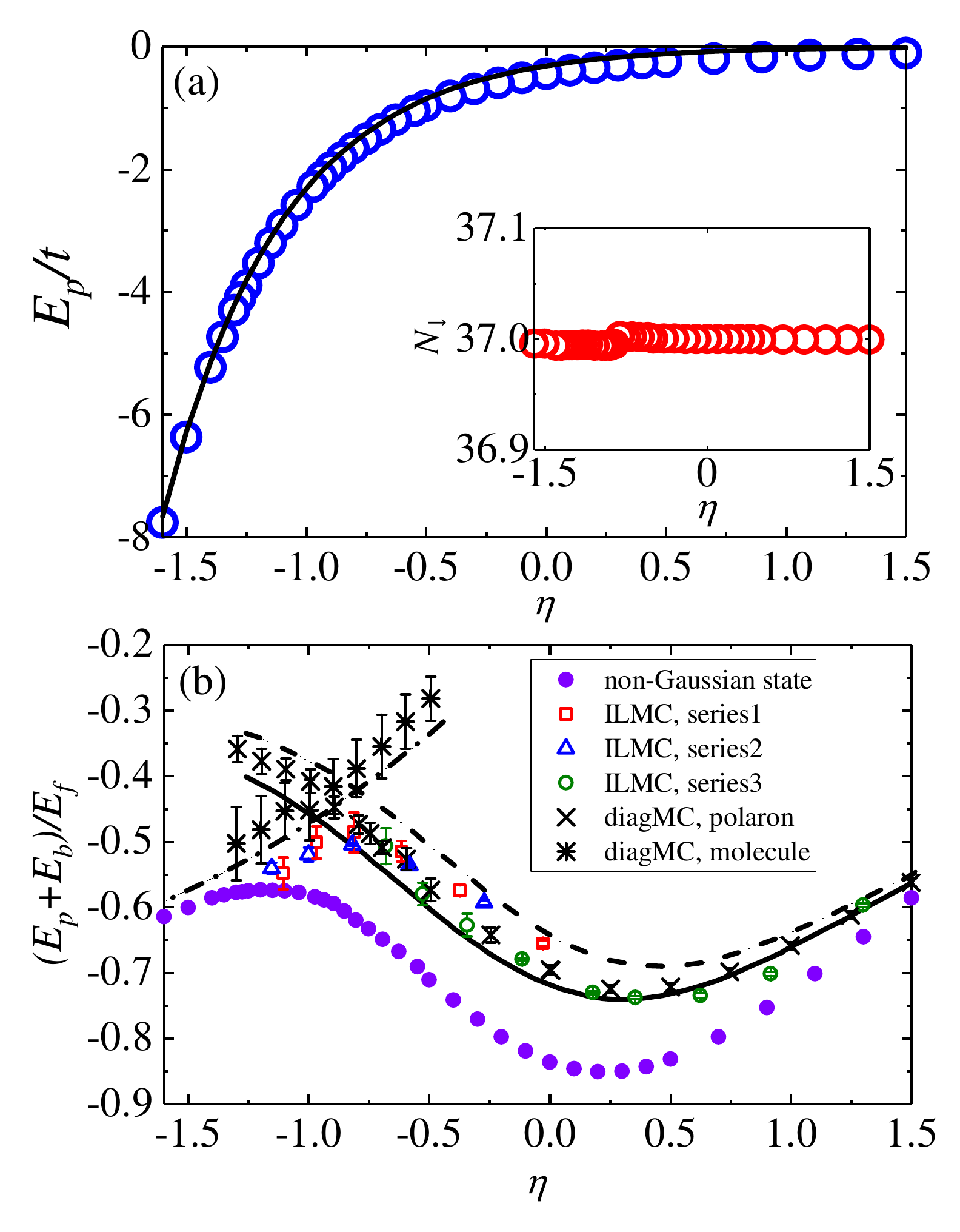}
	\end{center}
	\caption{\label{fig:2dEg}(Color online) (a) The non-Gaussian variational ground state energy (blue circles) of a 2D Fermi polaron with number of background particles $N_{\downarrow} \approx 37$ (inset) in a square lattice of size $50 \times 50$. The system is in the dilute limit with $N_\downarrow/\Omega \approx 0.015$. The energy saturates to the non-interacting value $E_p = 0$ in the weak coupling limit with large positive $\eta$, and to the two-body bound state energy $-E_b$ (black solid line) in the strong coupling limit with large negative $\eta$.
(b) A polaron--molecule evolution can be observed by plotting the polaron energy in a scaled way. Results are compared with outcome obtained using the Chevy-like polaron ansatz with one particle-hole pair excitation (dashed line)~\cite{Parish1}, the same ansatz with two particle-hole pair excitations (solid line)~\cite{Parish2}, the molecule variational wave function with one particle-hole excitation pairs (dashed-dotted line)~\cite{Parish1}. Some numerical solutions using diagrammatic quantum Monte Carlo (diagMC)~\cite{Vlietinck} and impurity lattice Monte Carlo (ILMC)~\cite{Bour} are also shown for comparison.}
\end{figure}
In a 2D continuum model, the interaction strength $g$ is characterized by the binding energy $E_b$ of a two-body bound state. To make a quantitative comparison, we first solve for the two-body bound state in the lattice Hamiltonian. In momentum space, the two-body Hamiltonian reads
\begin{eqnarray}
H^{(2)}=\sum_{ \mathbf{k},\sigma}\varepsilon'_{\mathbf{k}}c_{\mathbf{k}\sigma}^{\dag}c_{\mathbf{k}\sigma}
+\frac{g}{\Omega}\sum_{\mathbf{q},\mathbf{k},\mathbf{k'}}c_{\mathbf{q-k}\uparrow}^{\dag}c_{\mathbf{k}\downarrow}^{\dag}c_{\mathbf{k'}\downarrow}c_{\mathbf{q-k'}\uparrow}
\nonumber\\
\label{eqn:twobodyH}
\end{eqnarray}
with single particle dispersion
\begin{eqnarray}
\varepsilon'_{\mathbf{k}}=\varepsilon_{\mathbf{k}}+4=-2t\cos(k_{x})-2t\cos(k_{y})+4
\label{eqn:dispersion}
\end{eqnarray}
and number constraints $\sum\limits_{ \mathbf{k}}c_{\mathbf{k}\uparrow}^{\dag}c_{\mathbf{k}\uparrow}=\sum\limits_{ \mathbf{k}}c_{\mathbf{k}\downarrow}^{\dag}c_{\mathbf{k}\downarrow}=1$. Notice that we have shifted the zero energy point to the band bottom to get a direct comparison with the continuum model. The two-body wave function with zero total momentum can be generally written as
\begin{eqnarray}
|\Psi^{(2)}\rangle=\sum_{\mathbf{k}}\Psi^{(2)}_{\mathbf{k}}c_{\mathbf{k}\uparrow}^{\dag}c_{-\mathbf{k}\downarrow}^{\dag}|0\rangle.
\label{eqn:twobodyWave}
\end{eqnarray}
Substituting Eq.~(\ref{eqn:twobodyWave}) into the Schr\"{o}dinger equation
\begin{eqnarray}
H^{(2)}|\Psi^{(2)}\rangle=E^{(2)}|\Psi^{(2)}\rangle,
\end{eqnarray}
we obtain the following equation for the coefficients $\Psi_{\mathbf{k}}^{(2)}$
\begin{eqnarray}
2\varepsilon'_{\mathbf{k}}\Psi^{(2)}_{\mathbf{k}}+\frac{g}{\Omega}\sum\limits_{\mathbf{k'}}\Psi^{(2)}_{\mathbf{k'}}=E^{(2)}\Psi^{(2)}_{\mathbf{k}}.
\label{eqn:scheq}
\end{eqnarray}
Equation~(\ref{eqn:scheq}) leads to a self-consistent equation
%
\begin{eqnarray}
-\frac{1}{g}=\frac{1}{\Omega}\sum_{\mathbf{k}}\frac{1}{E_b+2\varepsilon'_{\mathbf{k}}},
\label{eqn:2body}
\end{eqnarray}
where $E_b=-E^{(2)}$ is the two-body binding energy. The two-body Hamiltonian Eq.~(\ref{eqn:twobodyH}) can also be solved numerically via the non-Gaussian variational method. In Fig.~\ref{fig:twobody} we show the results of $E_b$ obtained by the two methods, and find excellent agreement. This observation is another evidence for the validity of the variational approach.


With the connection between the lattice and continuum models built by Eq.~(\ref{eqn:2body}), we replace $g$ in the Hamiltonian Eq.~(\ref{eqn:H})
with $E_b$, and solve for the ground state with total momentum $\mathbf{K} = 0$. As in the 1D case, we define the polaron energy as the shift induced by interaction
\begin{eqnarray}
E_p=E(g) - E(g=0),
\end{eqnarray}
and plot the subtracted-scaled polaron energy $(E_p+E_b)/E_f$ versus the dimensionless interaction $\eta\equiv\frac{1}{2}\ln(2E_f/E_b)$ in Fig.~\ref{fig:2dEg}. Here, the Fermi energy $E_f=\varepsilon'(\mathbf{k}_f)$ is defined via the shifted dispersion relation Eq.~(\ref{eqn:dispersion}) with Fermi momentum $\mathbf{k}_f$. 

From Fig.~\ref{fig:2dEg}(a), we find that the variational result approaches to the value of a non-interaction system in the weak coupling limit with large positive $\eta$, and saturates to the two-body bound state energy $-E_b$ (solid line) in the strong coupling limit with large negative $\eta$. This observation suggests that the system transforms from a polaron to a molecule state by increasing the interaction from zero. In fact, by comparing with the energies of polaron and molecule states obtained by either Chevy-like ansatz or diagrammatic MC as shown in Fig.~\ref{fig:2dEg}(b), the results obtained by the non-Gaussian variational approach show good agreement in the corresponding weak and strong interacting limits. In the intermediate interaction regime, the non-Gaussian variational method reveals a fairly broad evolution from polaron to molecule states, with a ground state energy significantly lower than all other numerical and variational methods throughout the entire parameter region. We emphasize that in this calculation one does not need to assume any specific form of the trial wave function, and the results for different interaction strengths are obtained via the same algorithm with very economical numerical efforts. The numerical convergence is quite stable against different initial states and variational routes. From Fig.~\ref{fig:2dEg}(b), we estimate the polaron--molecule evolution takes place within the parameter region $-1.3 < \eta < -1$, which is approximately consistent with those obtained by the Chevy-like ansatz~\cite{Parish2} with $-0.97 < \eta < -0.80$, the diagMC method~\cite{Vlietinck} with $-1.1<\eta< -0.8$, and the ILMC method~\cite{Bour} with $-0.9 < \eta < -0.75$ for 2D systems, as well as the diagMC method with $-1.3< \eta<-0.9$ for quasi-2D geometries.~\cite{Kroiss} 

\section{Conclusion}
\label{sec:conclusion}

We study the polaron problem of Fermi Hubbard model in one- and two-dimensional square lattices. By employing the Lee-Low-Pines transformation to separate the impurity from the background fermions, and the Gaussian approximation for the resulting bath Hamiltonian, we obtain a variational wave function in the form of a non-Gaussian state. The ground state energy and other properties are obtained by solving the imaginary time evolution problem of the variational parameters. For one-dimensional lattices, we obtain the ground state energy and dispersion relation, and achieve excellent agreement with the matrix product states method. For the two-dimensional case, we focus on the dilute limit and find an evolution from the polaron to molecule states by varying the interaction strength, without assuming {\it in priori} any specific form of the state. The parameter region of the evolution is consistent with existing results obtained by variational method, diagrammatic quantum Monte Carlo simulation, and impurity lattice Monte Carlo algorithm. We emphasize that as the present method does not rely on the dimensionality or specific form of the lattice, it can be straightforwardly generalized to other lattice configurations in various dimensions.

\acknowledgments
This work is supported by the National Natural Science Foundation of China (Grant Nos. 11434011, 11522436, 11774425), the National Key R$\&$D Program of China (Grant No. 2018YFA0306501), the Beijing Natural Science Foundation (Grant No. Z180013), the Joint Fund of the Ministry of Education (Grant No. 6141A020333xx), and the Research Funds of Renmin University of China (GrantsNos. 16XNLQ03 and 18XNLQ15).


\end{document}